\documentclass[prd,superscriptaddress,nofootinbib]{revtex4}
\usepackage{amsmath,amssymb,epsfig,color}
\usepackage{natbib,ifthen}
\usepackage{bm}

\begin{document}

\newcommand{\fixme}[1]{{\textbf{Fixme: #1}}}
\newcommand{\AC}[1]{{\textcolor{red}{\textbf{AC}: #1}}}
\newcommand{\ACtwo}[1]{\textcolor{green}{\textbf{AC}: #1}}
\newcommand{\ABL}[1]{{\textcolor{blue}{\textbf{ABL}: #1}}}
\newcommand{\JC}[1]{{\textcolor{red}{\textbf{JC}: #1}}}

\newcommand{\detD}{{\det\!\cld}}
\newcommand{\clh}{\mathcal{H}}
\newcommand{\ud}{{\rm d}}
\renewcommand{\eprint}[1]{\href{http://arxiv.org/abs/#1}{#1}}
\newcommand{\adsurl}[1]{\href{#1}{ADS}}
\newcommand{\ISBN}[1]{\href{http://cosmologist.info/ISBN/#1}{ISBN: #1}}
\newcommand{\jcap}{J.\ Cosmol.\ Astropart.\ Phys.}
\newcommand{\sovast}{Soviet Astronomy}
\newcommand{\mnras}{Mon.\ Not.\ R.\ Astron.\ Soc.}
\newcommand{\progress}{Rep.\ Prog.\ Phys.}
\newcommand{\prlett}{Phys.\ Rev.\ Lett.}
\newcommand{\aap}{A\&A}
\newcommand{\aapr}{A\&A Rev.}
\newcommand{\vort}{\varpi}
\newcommand\ba{\begin{eqnarray}}
\newcommand\ea{\end{eqnarray}}
\newcommand\be{\begin{equation}}
\newcommand\ee{\end{equation}}
\newcommand\lagrange{{\cal L}}
\newcommand\cll{{\cal L}}
\newcommand\cln{{\cal N}}
\newcommand\clx{{\cal X}}
\newcommand\clz{{\cal Z}}
\newcommand\clv{{\cal V}}
\newcommand\cld{{\cal D}}
\newcommand\clt{{\cal T}}
\newcommand{\clw}{\mathcal{W}}

\newcommand{\ThreeJSymbol}[6]{\begin{pmatrix}
#1 & #3 & #5 \\
#2 & #4 & #6
 \end{pmatrix}}

\newcommand\clo{{\cal O}}
\newcommand{\cla}{{\cal A}}
\newcommand{\clp}{{\cal P}}
\newcommand{\clr}{{\cal R}}
\newcommand{\calW}{{\cal W}}
\newcommand{\uD}{{\mathrm{D}}}
\newcommand{\calE}{{\cal E}}
\newcommand{\calB}{{\cal B}}
\newcommand{\curl}{\,\mbox{curl}\,}
\newcommand\del{\nabla}
\newcommand\Tr{{\rm Tr}}
\newcommand\half{{\frac{1}{2}}}
\newcommand\fourth{{1\over 8}}
\newcommand\bibi{\bibitem}
\newcommand{\kf}{\beta}
\newcommand{\kfprod}{\alpha}
\newcommand\calS{{\cal S}}
\renewcommand\H{{\cal H}}
\newcommand\K{{\rm K}}
\newcommand\mK{{\rm mK}}
\newcommand\km{{\rm km}}
\newcommand\synch{\text{syn}}
\newcommand\opacity{\tau_c^{-1}}

\newcommand{\Psil}{\Psi_l}
\newcommand{\bsigma}{{\bar{\sigma}}}
\newcommand{\bI}{\bar{I}}
\newcommand{\bq}{\bar{q}}
\newcommand{\bv}{\bar{v}}
\renewcommand\P{{\cal P}}
\newcommand{\numfrac}[2]{{\textstyle \frac{#1}{#2}}}

\newcommand{\la}{\langle}
\newcommand{\ra}{\rangle}

\newcommand{\Omtot}{\Omega_{\mathrm{tot}}}
\newcommand\xx{\mbox{\boldmath $x$}}
\newcommand{\phpr} {\phi`}
\newcommand{\gam}{\gamma_{ij}}
\newcommand{\sqgam}{\sqrt{\gamma}}
\newcommand{\delk}{\Delta+3{\K}}
\newcommand{\dph}{\delta\phi}
\newcommand{\om} {\Omega}
\newcommand{\dom}{\delta^{(3)}\left(\Omega\right)}
\newcommand{\rar}{\rightarrow}
\newcommand{\Rar}{\Rightarrow}
\newcommand\gsim{ \lower .75ex \hbox{$\sim$} \llap{\raise .27ex \hbox{$>$}} }
\newcommand\lsim{ \lower .75ex \hbox{$\sim$} \llap{\raise .27ex \hbox{$<$}} }
\newcommand\bigdot[1] {\stackrel{\mbox{{\huge .}}}{#1}}
\newcommand\bigddot[1] {\stackrel{\mbox{{\huge ..}}}{#1}}
\newcommand{\Mpc}{\text{Mpc}}
\newcommand{\Al}{{A_l}}
\newcommand{\Bl}{{B_l}}
\newcommand{\eAl}{e^\Al}
\newcommand{\ix}{{(i)}}
\newcommand{\ixp}{{(i+1)}}
\renewcommand{\k}{\beta}
\newcommand{\HD}{\mathrm{D}}

\newcommand{\nonflat}[1]{#1}
\newcommand{\Cgl}{C_{\text{gl}}}
\newcommand{\Cgltwo}{C_{\text{gl},2}}
\newcommand{\He}{{\rm{He}}}
\newcommand{\Mhz}{{\rm MHz}}
\newcommand{\vx}{{\mathbf{x}}}
\newcommand{\ve}{{\mathbf{e}}}
\newcommand{\vv}{{\mathbf{v}}}
\newcommand{\vk}{{\mathbf{k}}}
\newcommand{\vl}{{\mathbf{l}}}
\newcommand{\vL}{\mathbf{L}}
\newcommand{\vn}{{\mathbf{n}}}
\newcommand{\valpha}{\boldsymbol{\alpha}}
\newcommand{\vnabla}{\boldsymbol{\nabla}}

\newcommand{\vnhat}{{\hat{\mathbf{n}}}}
\newcommand{\vkhat}{{\hat{\mathbf{k}}}}
\newcommand{\taueps}{{\tau_\epsilon}}

\newcommand{\vgrad}{{\mathbf{\nabla}}}
\newcommand{\fbarln}{\bar{f}_{,\ln\epsilon}(\epsilon)}


\title{Limitations of CMB $B$-mode template delensing}

\author{Antón Baleato Lizancos}
\email{a.baleatolizancos@ast.cam.ac.uk}
\affiliation{Institute of Astronomy and Kavli Institute for Cosmology, Madingley Road, Cambridge, CB3 0HA, UK}

\author{Anthony Challinor}
\email{a.d.challinor@ast.cam.ac.uk}
\affiliation{Institute of Astronomy and Kavli Institute for Cosmology, Madingley Road, Cambridge, CB3 0HA, UK}
 \affiliation{DAMTP, Centre
for Mathematical Sciences, Wilberforce Road, Cambridge CB3 0WA, UK}

\author{Julien Carron}
\email{julien.carron@unige.ch}
\affiliation{Universit\'e de Gen\`eve, D\'epartement de Physique Th\'eorique et CAP, 24 Quai Ansermet, CH-1211 Gen\`eve 4, Switzerland}



\begin{abstract}
Efforts to detect a primordial $B$-mode of CMB polarization generated by inflationary gravitational waves ought to mitigate the large variance associated with the $B$-modes produced by gravitational lensing, a process known as \emph{delensing}. A popular approach to delensing entails building a lensing $B$-mode template by mimicking the lensing operation, either at gradient order or non-perturbatively, using high-resolution $E$-mode observations and some proxy of the lensing potential. By explicitly calculating all contributions to two-loop order in lensing to the power spectrum of $B$-modes delensed with such a template in the noise-free limit, we are able to show that: (i) corrections to the leading-order calculation of the lensing $B$-mode power spectrum only enter at the $O(1)\,\%$ level because of extensive cancellations between large terms at next-to-leading order; (ii) these cancellations would disappear if a gradient-order template were to be built from unlensed or delensed $E$-modes, giving rise to a residual delensing floor of $O(10)\,\%$ of the original power; (iii) new cancellations arise when the lensed $E$-modes are used in the gradient-order template, allowing for the delensing floor to be as low as $O(1)\,\%$ of the original power in practical applications of this method; and (iv) these new cancellations would disappear for a non-perturbative template constructed from the lensed $E$-modes, reintroducing a residual delensing floor of $O(10)\,\%$. We further show that the gradient-order template outperforms the non-perturbative one in realistic scenarios with noisy estimates of the $E$-mode polarization and lensing potential. We therefore recommend that in practical applications of $B$-mode template delensing, where the template is constructed directly from the (filtered) observed $E$-modes, the gradient-order approach should be used rather than a non-perturbative remapping. 
\end{abstract}

\maketitle

\section{Introduction}

The wealth of cosmological data gathered in recent decades favours models where a period of accelerated expansion of space took place at very early times: \emph{cosmic inflation}. As a general and rather unique feature, inflationary models predict that a background of primordial gravitational waves would have been generated during that period along with fluctuations in the density. These primordial perturbations would have been present 380,000 years later, when the cosmic microwave background (CMB) was emitted, hence they are expected to have left an imprint in the temperature and polarization patterns of this relic light~\cite{Polnarev:1985,Kamionkowski:1996zd,Seljak:1996gy}. In fact, one can form a curl-like, $B$-mode component of polarization which, during recombination and in linear theory, is generated only by tensor fluctuations. Consequently, a detection of large-scale primordial $B$-modes would widely be considered direct evidence for cosmic inflation.

Unfortunately, the faint, primordial signal is obscured by $B$-modes generated from primordial $E$-modes by the process of gravitational lensing of the CMB as it propagates through the large-scale matter distribution of the Universe~\cite{Zaldarriaga:1998ar}. On large angular scales, where the power spectrum of primordial $B$-modes peaks ($l<200$), the lensing power spectrum resembles that of white noise, comparable in amplitude to the noise power of an experiment with a sensitivity of $\Delta_{P}=5\,\mu \text{K\,arcmin}$, approximately that of experiments coming online at the time of writing. Lensing-induced $B$-modes are now routinely observed in high-precision polarization surveys, following the first detection (with data from the South Pole Telescope) by Ref.~\cite{ref:hanson_13}. In order to be able to detect any small, primordial $B$-mode signal, the variance associated with the lensing contribution ought to be mitigated --- a procedure known as \emph{delensing}, which several groups have already applied to real data~\cite{Carron:2017vfg, ref:spt_17, Planck2018:lensing, ref:polarbear_delensing_19, ref:han_20}.

Since the effect of lensing on the CMB is very well approximated as a re-mapping of the unlensed anisotropies by the gradient of the lensing potential, extensive delensing can be achieved by reversing these deflections. For $B$-modes, this re-mapping approximation should be valid until polarization sensitivities become better than $O(0.01)\,\mu\text{K\,arcmin}$, corresponding to lensing residuals $O(10^{-4})$ of the original power, at which point non-remapping effects such as rotation of the emission-angle~\cite{ref:lewis_17}, apparent warping of the last-scattering surface due to Shapiro time-delays~\cite{ref:hu_and_cooray,ref:lewis_17} and post-Born field rotation~\cite{Hirata:2003ka, ref:pratten_16} (dominant only on small angular scales) will present a source of $B$-mode noise not reducible by standard delensing. We shall neglect non-remapping corrections for the remainder of this work.

For $B$-mode delensing, the information that is needed to delens the large angular scales of interest is contained in the intermediate and small-scale fluctuations~\cite{ref:smith_12_external}. Given that it is challenging to make stable large-scale measurements with the large-aperture telescopes needed to access the small-scale fluctuations, upcoming ground-based observatories will feature one or more large-aperture telescopes (for lensing and other CMB science needing high angular resolution) and a set of small-aperture telescopes targeting the primordial $B$-mode signal on large angular scales~\cite{CMBS4:2019RefDesign,SO:2019Forecast}. The two surveys will have different footprints, with the small-aperture survey contained within the wider large-aperture one. In this set-up, a convenient way to perform delensing is to form a ``template'' estimating the particular realization of lensing $B$-modes present on the sky by combining high-resolution $E$-mode observations with some proxy of the lensing potential, and to subtract this template from the $B$-modes observed by the small-aperture telescopes. The template may be constructed using a linear expansion in the lensing deflections (which we refer to as a gradient-order template) or using a non-perturbative remapping scheme. Both approaches have already been successfully applied to real data: Ref.~\cite{ref:spt_17} used a leading-order template for a first demonstration of $B$-mode delensing, while Refs~\cite{Planck2018:lensing,ref:polarbear_delensing_19} used non-perturbative templates for their similarly successful analyses. The gradient-order method has been particularly popular in the literature for forecasts of performance (e.g., Refs~\cite{ref:marian_07, ref:cmbpol, ref:smith_12_external, ref:namikawa_14, ref:namikawa_15, ref:simard_15, ref:sherwin_15, ref:S4_forecast, ref:errard_16, ref:namikawa_16, ref:yu_17,ref:core, ref:manzotti_18, ref:karkare_19}) and characterizations of systematic effects (e.g., Refs~\cite{ref:namikawa_17, ref:namikawa_19, ref:beck_20, ref:baleato_20_internal, ref:cib_delensing_biases}) owing, perhaps, to its analytic transparency and ease of computation. A further reason for the ubiquity of the gradient-order template method is that it is assumed to track the true lensing $B$-modes very accurately.

In this work, we put the last statement to the test, aiming to quantify and understand the intrinsic limitations on the residual $B$-mode power after delensing with gradient-order and non-perturbative templates. We begin, in Sec.~\ref{sec:template}, with a brief introduction to template delensing. Then, in Secs.~\ref{sec:delensed_ps} and~\ref{sec:delensed_psnp}
we calculate the power spectrum of delensed $B$-modes to $O(\phi^4)$ (i.e., two-loop order in the power spectrum of the lensing potential, $\phi$) for the gradient-order and non-perturbative templates, respectively. We demonstrate that the former has better performance due to cancellations between the lensing corrections to the $E$-modes used in the template, and the $O(\phi^2)$ contributions to the lensed $B$-modes, which do not arise in the non-perturbative case. We also point out similar calculations in the lensed $B$-mode power spectrum itself at $O(\phi^4)$ in Sec.~\ref{sec:delensed_ps}. Much of our discussion concerns the fundamental limitations of template methods, which we illustrate by considering the idealised case of noise-free $E$-mode measurements and access to the true lensing potential. However, we show in Sec.~\ref{sec:delensed_psnp} that our conclusion regarding the relative performance of gradient-order and non-perturbative templates still holds, albeit with more marginal differences, in the practical case of noisy observations and an imperfectly correlated lensing proxy.

\section{Template delensing}
\label{sec:template}

We work in the flat-sky limit throughout for simplicity.
While differences do exist at the percent level between flat-sky and spherical results, for example, in the lensed $B$-mode power spectrum~\cite{Challinor:2005jy}, our findings on the limitations of template delensing should hold similarly in the spherical case.
We follow the notation of~\cite{Lewis:2006fu}, so that the Stokes parameters defined on the global $x$--$y$ basis are related to the $E$- and $B$-modes in Fourier space as
\begin{equation}
(Q\pm i U)(\vx) = - \int \frac{d^2 \vl}{2\pi} \, \left[E(\vl)\pm i
  B(\vl)\right] e^{\pm 2i\psi_{\vl}} e^{i \vl\cdot\vx} \, ,
\end{equation}
where $\psi_{\vl}$ is the angle between $\vl$ and the
$x$-direction. Ignoring primordial $B$-modes, the $B$-modes generated
from $E$-modes by lensing are, up to second order in lensing
displacements,
\begin{multline}
\tilde{B}(\vl) = - \int \frac{d^2 \vl_1}{2\pi}\, \sin 2(\psi_{\vl_1} -
\psi_{\vl}) \vl_1 \cdot (\vl-\vl_1) E(\vl_1) \phi(\vl-\vl_1) \\
+ \frac{1}{2} \int \frac{d^2 \vl_1}{2\pi} \int \frac{d^2 \vl_2}{2\pi}
\sin 2(\psi_{\vl_1}-\psi_{\vl}) \vl_1 \cdot \vl_2 \vl_1 \cdot (\vl -
\vl_1 - \vl_2) E(\vl_1) \phi(\vl_2) \phi(\vl-\vl_1-\vl_2) + \cdots \,
.
\label{eq:lensedB}
\end{multline}
Here, $\phi$ is the lensing potential so the lensing displacements are
$\valpha = \vnabla \phi$. We denote lensed fields with a tilde. In particular, we write the contribution to the lensed
$B$-mode at $n$th order in $\phi$ as $\tilde{B}^{(n)}$. Further
introducing the linear functional $\calB_{\vl}[P]$, which extracts the
$B$-modes at $\vl$ from the real-space polarization field $P=Q+iU$, we
can write
\begin{equation}
\tilde{B}^{(n)}(\vl) = \frac{1}{n!} \calB_{\vl}[\alpha^{i_1} \ldots
\alpha^{i_n} \nabla_{i_1} \ldots \nabla_{i_n} P^E] \, ,
\label{eq:functional}
\end{equation}
where $P^E$ is the unlensed polarization field constructed from
$E(\vl)$.

Given the observed, noisy $E$-mode polarization after
beam-deconvolution, $E^{\text{obs}}(\vl)$, and some proxy for the
lensing potential $\phi^{\text{proxy}}$, we can form a leading-order
  template (which we shall refer to as the gradient-order template) for the lens-induced $B$-modes:
\begin{equation}
\tilde{B}^{\text{temp}}(\vl) = - \int \frac{d^2 \vl_1}{2\pi}\, \sin 2(\psi_{\vl_1} -
\psi_{\vl}) \vl_1 \cdot (\vl-\vl_1) \calW^E_{l_1}
E^{\text{obs}}(\vl_1)
\calW^\psi_{|\vl-\vl_1|}\phi^{\text{proxy}}(\vl-\vl_1) \, .
\label{eq:template}
\end{equation}
Here, the Wiener filters
\begin{equation}
\calW^E_l \equiv \frac{\tilde{C}_l^{EE}}{C_l^{EE,\text{tot}}} \quad
\text{and} \quad 
\calW^{\phi}_l \equiv \frac{C_l^{\phi
    \phi_{\text{proxy}}}}{C_l^{\phi_{\text{proxy}}\phi_{\text{proxy}}}}
\, ,
\label{eq:wienerfilters}
\end{equation}
are chosen to minimise the residual power in
$\tilde{B}(\vl) - \tilde{B}^{\text{temp}}(\vl)$ at leading
order (note that $C_l^{EE,\text{tot}}$ includes noise). A template that is non-perturbative in the lens remapping can also be constructed:
\begin{equation}
\tilde{B}^{\text{temp}}_{\text{non-pert}}(\vl) = \calB_{\vl} \left(
P^{\mathcal{W}^E \ast E^{\text{obs}}}\left[\vx + \vnabla \left(\mathcal{W}^\phi \ast \phi^{\text{proxy}}\right)\right]\right) \, ,
\label{eq:template_np}
\end{equation}
where, for example, $\mathcal{W}^E \ast E^{\text{obs}}$ are the Wiener-filtered $E$-modes. Taylor expanding at gradient order recovers the gradient-order template in Eq.~\eqref{eq:template}.

For most of this work, we shall assume that the polarization measurements
are noise-free and that we have access to the true lensing potential,
in which case both Wiener filters in Eq.~\eqref{eq:wienerfilters} equal one. However, we relax these assumptions in Sec.~\ref{sec:delensed_psnp}, where we compare the performance of gradient-order and non-perturbative templates in a more practical context.

\section{Delensing with a gradient-order template}
\label{sec:delensed_ps}
It is sometimes argued, e.g., Ref.~\cite{ref:spt_17}, that
delensing with a gradient-order template constructed from ideal
(lensed) $E$-modes and the true $\phi$ is very
accurate since the non-perturbative lensed $B$-mode power differs only
at the percent level from the power spectrum of
$\tilde{B}^{(1)}$~\cite{Challinor:2005jy}. As we shall see, the
conclusion is correct that the residual power after template delensing
can be at the percent level. However, the logic above is somewhat
flawed since:
\begin{enumerate}
\item the template is constructed from the lensed $E$-modes, not the
  unobservable unlensed ones; and
\item the power spectrum of the residuals after template delensing is
  not simply related to the difference between the power spectra of
  the template and $\tilde{B}(\vl)$. 
\end{enumerate}

To see some of the subtleties, consider forming a gradient-order template with the
unlensed $E$-modes, which might seem a desirable thing to be able to
do. Of course, if we made a non-perturbative template this would be
perfect, but we are interested in this section in a gradient-order template of the
form in Eq.~\eqref{eq:template}. If we subtract this template, the
residual $B$-modes are $\tilde{B}^{(2)}(\vl) + \tilde{B}^{(3)}(\vl) +
\cdots$. \emph{The power spectrum of these residuals is much larger
  than the percent-level difference between the power spectra of
  the template and $\tilde{B}(\vl)$}. 

\begin{figure}
\centering
\includegraphics[width=0.8\textwidth]{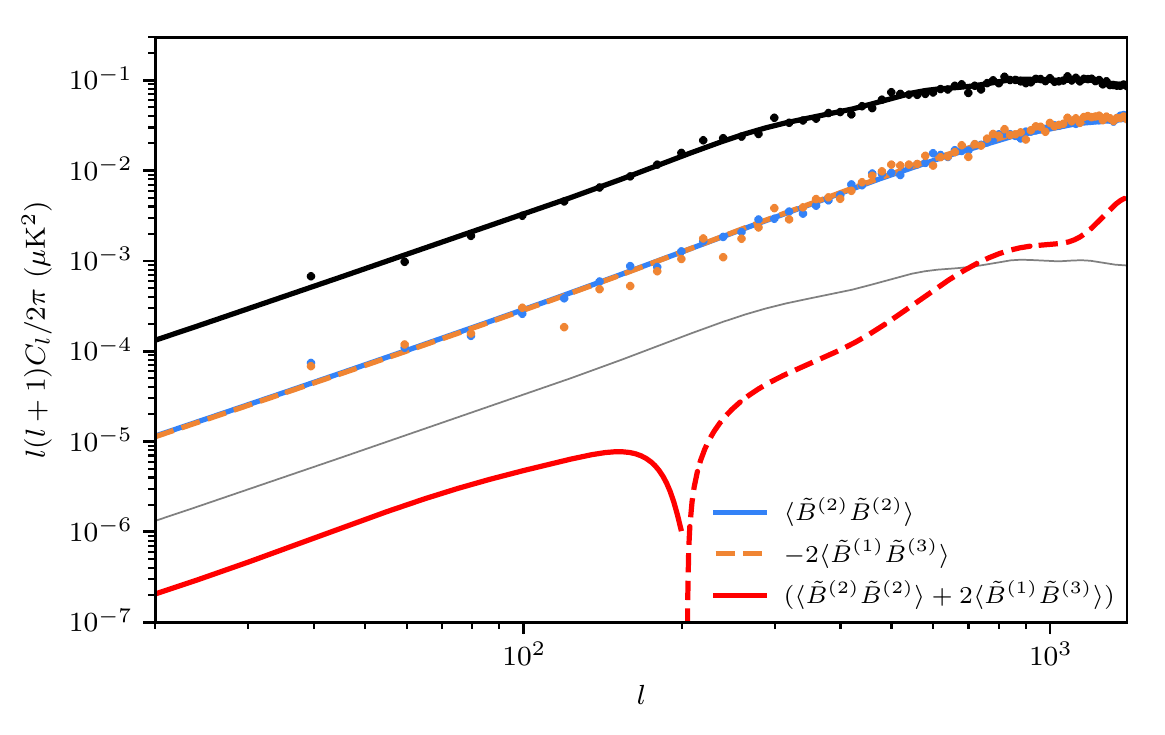}
\caption{Power spectrum of $\tilde{B}^{(1)}(\vl)$ estimated from a single flat-sky simulation (black points) compared to the non-perturbative, spherical calculation of the lens-induced $B$-mode power from \textsc{CAMB}~\cite{Challinor:2005jy,Lewis:1999bs} (black solid). The leading-order difference between these spectra involves the sum of $\langle \tilde{B}^{(2)}(\vl)
\tilde{B}^{(2)}(\vl') \rangle'$ (blue) and $2\langle \tilde{B}^{(1)}(\vl) \tilde{B}^{(3)}(\vl')
\rangle'$ (with minus this shown in dashed, orange). Separately, these
two spectra are large in magnitude, of $O(10)\,\%$ of the lensed $B$-mode
spectrum on large scales, but they cancel rather precisely to leave
only sub-percent-level corrections to the lensed $B$-mode power spectrum (red solid for positive values, red dashed for negative). For reference, the thin grey line shows the non-perturbative lensing power spectrum, scaled to $1\%$ of its original amplitude. Analytic evaluations are shown as solid lines, and estimates derived from a single flat-sky simulation as points.}
\label{fig:twotwo_onethree_cancel}
\end{figure}

The difference between the power spectra of
  the template (constructed at gradient order with the unlensed $E$-modes) and of $\tilde{B}(\vl)$ at second order in $C_l^{\phi\phi}$ is of the form
\begin{equation}
\Delta C_l^{BB} = 2 \langle \tilde{B}^{(1)}(\vl) \tilde{B}^{(3)}(\vl')
\rangle' + \langle \tilde{B}^{(2)}(\vl) \tilde{B}^{(2)}(\vl')
\rangle' \, ,
\label{eq:deltapowerunl}
\end{equation}
where the primes on the expectation values denote that the
delta-functions $\delta^{(2)}(\vl+\vl')$ are removed. This is simply the contribution to the lensed $B$-mode power at second order in $C_l^{\phi\phi}$. However, the power
spectrum of the residual $B$-modes after gradient-order template delensing with the
unlensed $E$-modes is
\begin{equation}
C_l^{BB,\text{resid}} = \langle \tilde{B}^{(2)}(\vl) \tilde{B}^{(2)}(\vl')
\rangle' 
\end{equation}
to the same order in $C_l^{\phi\phi}$.
There turns out to be a strong cancellation between the two terms on
the right-hand side of Eq.~\eqref{eq:deltapowerunl}, each of which are
separately at the $O(10)\,\%$ level of the lensed $B$-mode power; see Fig.~\ref{fig:twotwo_onethree_cancel}.

We first calculate $\langle \tilde{B}^{(2)}(\vl)
\tilde{B}^{(2)}(\vl') \rangle'$ from Eq.~\eqref{eq:lensedB}; we find
\begin{equation}
\langle \tilde{B}^{(2)}(\vl)
\tilde{B}^{(2)}(\vl') \rangle' = \frac{1}{2} \int \frac{d^2
  \vl_1}{(2\pi)^2} \int \frac{d^2 \vl_2}{(2\pi)^2} \, \sin^2
2(\psi_{\vl_1} - \psi_{\vl} ) (\vl_1\cdot \vl_2)^2 \left[\vl_1\cdot
  (\vl-\vl_1-\vl_2)\right]^2 C_{l_1}^{EE} C_{l_2}^{\phi\phi}
C_{|\vl-\vl_1 - \vl_2|}^{\phi\phi} \, .
\label{eq:twotwo}
\end{equation}
For $\langle \tilde{B}^{(1)}(\vl) \tilde{B}^{(3)}(\vl')
\rangle'$, it is simplest to make use of Eq.~\eqref{eq:functional}, so
that
\begin{align}
\langle \tilde{B}^{(1)}(\vl) \tilde{B}^{(3)}(\vl')
\rangle' &= \frac{1}{3!} \langle \calB_{\vl}[\valpha \cdot \vnabla
           P^E] \calB_{\vl'}[\alpha^{i_1} \alpha^{i_2} \alpha^{i_3}
           \nabla_{i_1} \nabla_{i_2} \nabla_{i_3} P^E]\rangle '
           \nonumber \\
&= \frac{3}{2\times 3!} \langle \valpha^2 \rangle \langle \calB_{\vl}[\valpha \cdot \vnabla
           P^E] \calB_{\vl'}[\valpha \cdot \vnabla
           \nabla^2 P^E]\rangle ' \, ,
\end{align}
where we have used $\langle \alpha_i \alpha_j \rangle = \delta_{ij} \langle
\valpha^2 \rangle /2$ at a point. It follows that the calculation of
this ``1--3'' term is simply related to the power spectrum of
$\tilde{B}^{(1)}$, and we have
\begin{equation}
2 \langle \tilde{B}^{(1)}(\vl) \tilde{B}^{(3)}(\vl')
\rangle' = - \frac{\langle \valpha^2 \rangle}{2} \int \frac{d^2 \vl_1}{(2\pi)^2} \, \sin^2
2(\psi_{\vl_1} - \psi_{\vl}) \left[\vl_1 \cdot (\vl-\vl_1)\right]^2
l_1^2 C_{l_1}^{EE} C_{|\vl-\vl_1|}^{\phi\phi} \, .
\label{eq:onethree}
\end{equation}

On large scales, the integral on the right of Eq.~\eqref{eq:onethree} is dominated by small-scale $E$-modes and similarly small-scale lenses, as for the power spectrum of $\tilde{B}^{(1)}(\vl)$, making $\langle \tilde{B}^{(1)}(\vl) \tilde{B}^{(3)}(\vl')\rangle$ approximately constant. (Indeed, the additional factor of $l_1^2$ in the integrand compared to the power spectrum of $\tilde{B}^{(1)}(\vl)$ accentuates this coupling to smaller-scale lenses and $E$-modes.) On the other hand, the mean-squared deflection angle $\langle \valpha^2 \rangle$ preferentially receives contributions from degree-scale lenses (i.e., the coherence length of the deflections). Physically, the large-scale modes of $\tilde{B}^{(3)}$ that correlate with the large-scale modes of $\tilde{B}^{(1)}$ are mostly sourced by the action of one small-scale lens and two larger-scale lenses on the small-scale (unlensed) $E$-modes. Roughly, this can be thought of as the small-scale lens linearly producing $B$-modes from the small-scale $E$-modes, and then these being displaced rigidly by lenses on larger scales, preserving the $B$-mode character.

A similar thing happens for $\tilde{B}^{(2)}$: the dominant contribution on large scales is from one small-scale lens acting on small-scale unlensed $E$-modes, followed by displacement by one large-scale lens. Equivalently, the integral in Eq.~\eqref{eq:twotwo} is dominated on large scales by modes with $l \ll l_1$ and $l_2 \ll l_1$, with an equal contribution (by symmetry of the integrand) from the disjoint region with $\vl_2 \rightarrow \vl - \vl_1 - \vl_2$. The two contributions correspond to the two orderings of the large- and small-scale lenses in $\tilde{B}^{(2)}(\vl) =\calB_{\vl}[\alpha^{i}\alpha^{j} \nabla_{i} \nabla_{j} P^E]/2$. Evaluating the integral in these limits, we find
\begin{align}
\langle \tilde{B}^{(2)}(\vl)
\tilde{B}^{(2)}(\vl') \rangle'
&\approx \frac{1}{2} \left(\int \frac{d^2
  \vl_2}{(2\pi)^2} \, l_2^2 C_{l_2}^{\phi\phi} \right) 
\int \frac{d^2 \vl_1}{(2\pi)^2} \, \sin^2
2(\psi_{\vl_1} - \psi_{\vl}) 
l_1^6 C_{l_1}^{EE} C_{l_1}^{\phi\phi} \, , \nonumber \\
&\approx \frac{\langle \valpha^2 \rangle}{2} \frac{1}{4\pi}\int \frac{dl_1}{l_1}
l_1^4 C_{l_1}^{\phi\phi} l_1^4 C_{l_1}^{EE} \, ,
\label{eq:twotwolargescaleapprox}
\end{align}
which is independent of $l$. Here, we have used 
\begin{equation}
\langle \valpha^2 \rangle = \int \frac{d^2
  \vl}{(2\pi)^2} \, l^2 C_{l}^{\phi\phi} \, .
\end{equation}
Evaluating Eq.~\eqref{eq:onethree} in the large-scale limit, we see that
\begin{equation}
\langle \tilde{B}^{(2)}(\vl)
\tilde{B}^{(2)}(\vl') \rangle' \approx - 2 \langle \tilde{B}^{(1)}(\vl) \tilde{B}^{(3)}(\vl')
\rangle' \, ,
\end{equation}
and so the two terms approximately cancel in Eq.~\eqref{eq:deltapowerunl}. This result is validated numerically in Fig.~\ref{fig:twotwo_onethree_cancel}, showing that the $O(\phi^4)$ contribution to the lensed $B$-mode spectrum is below 1\,\% of the total lensing power, despite the power spectrum of $\tilde{B}^{(2)}$ being around 10\,\% of the lensing power.

The approximate cancellation between the ``1--3'' and ``2--2'' terms arises since displacing the $B$-modes, which are produced by the action of the small-scale lens on the small-scale unlensed $E$-modes, with large-scale lensing deflections has a relatively minor effect on their power spectrum due to statistical anisotropy (only relative displacements matter). However, there can still be significant changes in the $B$-mode fields themselves if the size of the displacements are comparable to the coherence scale of the field being displaced. In this way, $\tilde{B}^{(2)}$ can have power that is a sizable fraction of $\tilde{B}^{(1)}$ (i.e., around 10\,\% here), but the correction to the total lensed $B$-mode power at second order in $C_l^{\phi\phi}$ remains small (around 1\,\%). Similar cancellations are also the reason why the lensing correction to the temperature power spectrum remains small on the intermediate scales of the acoustic peaks.

We now consider the physically more relevant case of delensing with a gradient-order template constructed with the \emph{lensed} $E$-modes.
Fortunately, cancellations also occur in this case, leaving the power spectrum of the delensed $B$-modes at the 1\,\% level of the lensed $B$-mode power. Here, the dominant cancellation
is between $\tilde{B}^{(2)}$ and the contribution to the template of the leading-order
change in the $E$-modes due to lensing (although we verify using simulations that similar cancellations also occur at higher orders). In
Ref.~\cite{Pan:2017trj}, it was argued that this cancellation is
responsible for the lack of bias seen in their recovery of
the tensor-to-scalar ratio $r$ across simulations of the delensing process,
but no quantitative details were given. Here, we fill in these details.

\begin{figure}
\centering
\includegraphics[width=0.87\textwidth]{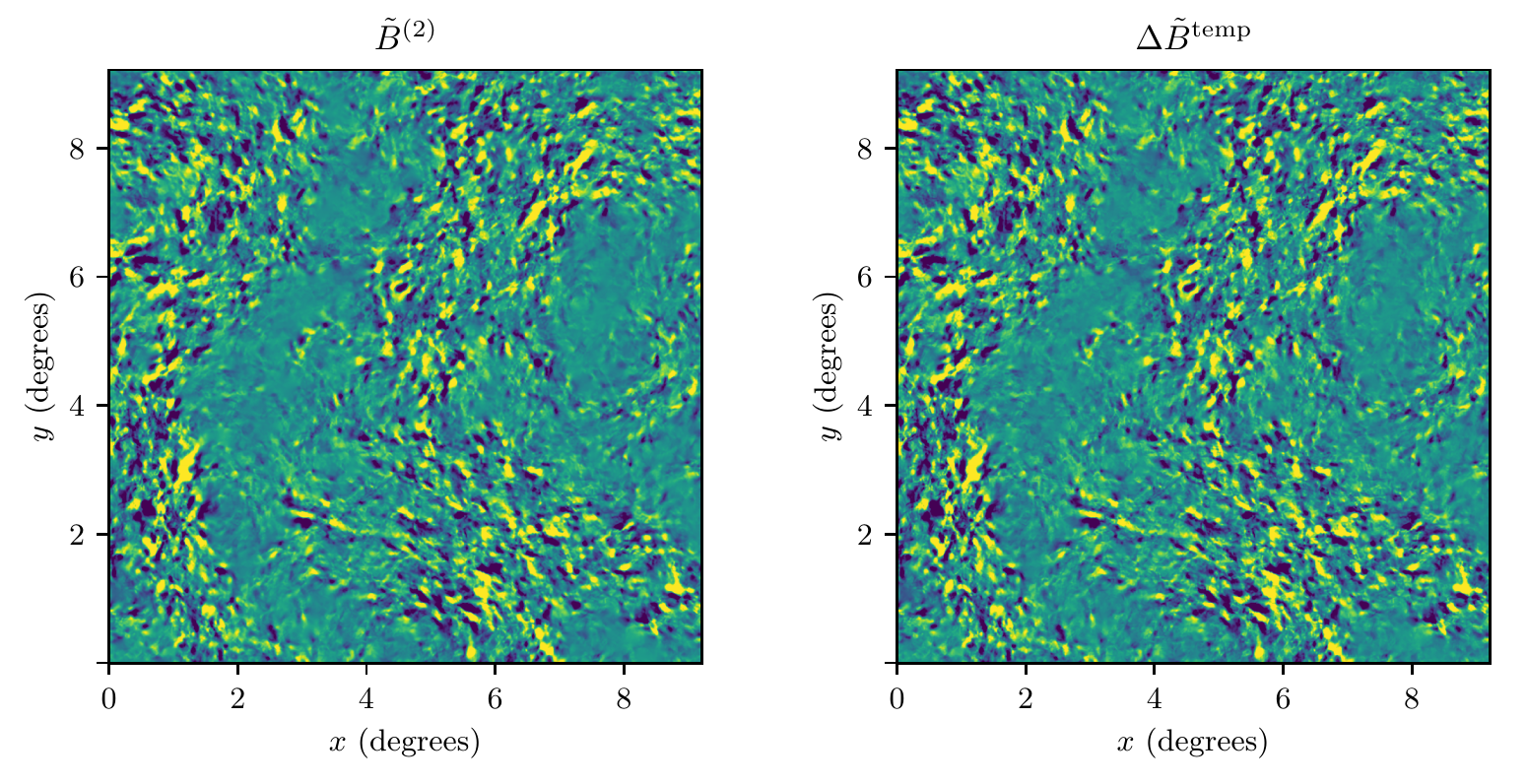}
\caption{Real-space scalar fields derived from $\tilde{B}^{(2)}(\vl)$
  (left) and the contribution to the gradient-order $B$-mode template from the linear
  lensing correction to $E$-modes, $\Delta \tilde{B}^{\text{temp}}(\vl)$
  (right). The plotted intensity ranges from $-0.5\,\mu\text{K}$ (dark blue) to $0.5\,\mu\text{K}$ (yellow). The similarity of these fields and their non-Gaussian nature are clearly apparent.}
\label{fig:compare_lensB2_lensE1convphi}
\end{figure}

If we construct the $B$-mode lensing template at gradient order using the lensed
$E$-modes, the leading-order correction to the template is
\begin{equation}
\Delta \tilde{B}^{\text{temp}}(\vl) = - \int \frac{d^2 \vl_1}{2\pi}\, \sin 2(\psi_{\vl_1} -
\psi_{\vl}) \vl_1 \cdot (\vl-\vl_1) \tilde{E}^{(1)}(\vl_1) \phi(\vl-\vl_1) \, ,
\label{eq:templatecorr}
\end{equation}
where $\tilde{E}^{(1)}(\vl)$ is the first-order change in $E$-modes
due to lensing:
\begin{align}
\tilde{E}^{(1)}(\vl) &= \calE_{\vl}[\valpha\cdot\vnabla P^E] \nonumber \\
&= - \int \frac{d^2 \vl_1}{2\pi}\, \cos 2(\psi_{\vl_1} -
\psi_{\vl}) \vl_1 \cdot (\vl-\vl_1) E(\vl_1) \phi(\vl-\vl_1) \, ,
\end{align}
where $\calE_\vl[P]$ extracts the $E$-modes at wavevector $\vl$ from
$P$. The residual $B$-modes after delensing with such a template become
$\tilde{B}^{(2)}(\vl) - \Delta \tilde{B}^{\text{temp}}(\vl)$ to second
order in $\phi$. A simulated realization of the fields $\tilde{B}^{(2)}(\vl)$
and $\Delta \tilde{B}^{\text{temp}}(\vl)$ are shown in
Fig.~\ref{fig:compare_lensB2_lensE1convphi}. The fields are clearly
very similar and so we expect the residual $B$-mode power after
delensing to be much smaller than the power of either field alone
(which, recall, for $\tilde{B}^{(2)}(\vl)$ is $O(10)\,\%$ of the power
of $\tilde{B}^{(1)}(\vl)$ on large scales). To understand the similarity of
$\tilde{B}^{(2)}(\vl)$ and $\Delta \tilde{B}^{\text{temp}}(\vl)$, we note that
\begin{align}
B^{(2)}(\vl) &= \frac{1}{2} \calB_{\vl} [\alpha^i \alpha^j \nabla_i \nabla_j P^E] \nonumber \\
&\approx \calB_{\vl} [\alpha^i_{\text{short}} \alpha^j_{\text{long}} 
\nabla_i \nabla_j P^E] \quad \text{(large scales)} \, ,
\end{align}
where the approximation, valid on large scales, is that $\tilde{B}^{(2)}$ is dominated by a large-scale deflection, $\valpha_{\text{long}}$, and a small-scale deflection, $\valpha_{\text{short}}$, as discussed above. For the leading-order correction to the template from the lensed $E$-modes, we have
\begin{align}
\Delta B^{\text{temp}}(\vl) &= \calB_\vl\left[\alpha^i \nabla_i P^{\tilde{E}^{(1)}}\right] \nonumber \\
&\approx \calB_\vl\left[\alpha^i \nabla_i \left(\alpha^j_{\text{long}} \nabla_j P^{E}\right)\right] \nonumber \\
&\approx \calB_\vl\left[\alpha_{\text{short}}^i \nabla_i \left(\alpha^j_{\text{long}} \nabla_j P^{E}\right)\right] \quad \text{(large scales)}\nonumber \\
&\approx \calB_\vl\left[\alpha_{\text{short}}^i \alpha_{\text{long}}^j \nabla_i \nabla_j P^{E}\right] \, .
\label{eq:DeltaBtempapprox}
\end{align}
Here, the approximation in the second line ignores the $B$-modes in $\valpha \cdot \vnabla P^E$, i.e.,
\begin{align}
P^{\tilde{E}^{(1)}} = \valpha \cdot \vnabla P^E - P^{\tilde{B}^{(1)}} \approx \valpha \cdot \vnabla P^E \,,
\end{align}
which is valid since the power spectrum of $\tilde{B}^{(1)}$ is only $O(1)\,\%$ of the power spectrum of $\tilde{E}^{(1)}$ on the small scales ($l \sim 1500$) where the power in the latter peaks. In the second and third lines of Eq.~\eqref{eq:DeltaBtempapprox}, we have used the fact that it is mostly large-scale (i.e., degree-scale) lenses that contribute to $\tilde{E}^{(1)}$ while smaller-scale lenses displace this to produce large-scale $B$-modes. In the final line, we have ignored the derivative of $\valpha_{\text{long}}$ compared to that of the small-scale $\vnabla P^E$. With these approximations, we see that $\Delta B^{\text{temp}}(\vl) \approx \tilde{B}^{(2)}(\vl)$.

We now calculate the power spectrum of residual $B$-modes after delensing, with a gradient-order template formed using lensed $E$-modes, correct to second order in $C_l^{\phi\phi}$:
\begin{equation}
C_l^{BB,\text{resid}} = \langle \tilde{B}^{(2)}(\vl) \tilde{B}^{(2)}(\vl') \rangle' - 2  \langle \tilde{B}^{(2)}(\vl) \Delta \tilde{B}^{\text{temp}}(\vl') \rangle' + \langle \Delta \tilde{B}^{\text{temp}}(\vl) \Delta \tilde{B}^{\text{temp}}(\vl') \rangle'\, .
\label{eq:BBresidualpower}
\end{equation}
The first of these terms was already calculated in Eq.~\eqref{eq:twotwo}. On the other hand, the cross-term is given by
\begin{align}
\langle \tilde{B}^{(2)}(\vl) \Delta \tilde{B}^{\text{temp}}(\vl') \rangle' =  \int \frac{d^2 \vl_1}{(2\pi)^2} \frac{d^2 \vl_2}{(2\pi)^2} &  \sin 2\left(\psi_{\vl_2}-\psi_{\vl}\right) \sin 2\left(\psi_{\vl_1}-\psi_{\vl}\right) \cos 2\left(\psi_{\vl_1}-\psi_{\vl_2}\right) \nonumber\\
& \times \left[ \left(\vl_2 -\vl \right) \cdot \vl_1 \right] \left[ \left(\vl_2 -\vl \right) \cdot \vl_2 \right] \left[ \left(\vl_2 -\vl_1 \right) \cdot \vl_1 \right]^2  C_{l_1}^{EE} C_{|\vl - \vl_2|}^{\phi\phi} C_{|\vl_2 - \vl_1|}^{\phi\phi},
\label{eq:B2xtemp}
\end{align}
and the last term, $\langle \Delta \tilde{B}^{\text{temp}}(\vl) \Delta \tilde{B}^{\text{temp}}(\vl') \rangle'$, receives contributions from the following two couplings:
\begin{align}
U_l = \int \frac{d^2 \vl_1}{(2\pi)^2} \frac{d^2 \vl_2}{(2\pi)^2} \sin^2 2\left(\psi_{\vl_2}-\psi_{\vl}\right) \cos^2 2\left(\psi_{\vl_1}-\psi_{\vl_2}\right) [\vl_2 \cdot (\vl - \vl_2) \vl_1 \cdot (\vl_2 - \vl_1)]^2 C_{l_1}^{EE} C_{|\vl - \vl_2|}^{\phi\phi} C_{|\vl_2 - \vl_1|}^{\phi\phi} \, ;
\label{eq:tempxtempU}
\end{align}
and
\begin{align}
V_l = \int \frac{d^2 \vl_1}{(2\pi)^2} \frac{d^2 \vl_2}{(2\pi)^2} & \sin 2\left(\psi_{\vl_2}-\psi_{\vl}\right) \cos 2\left(\psi_{\vl_1}-\psi_{\vl_2}\right) \sin 2\left(\psi_{\vl - \vl_2 + \vl_1}-\psi_{\vl}\right) \cos 2\left(\psi_{\vl_1}-\psi_{\vl-\vl_2 + \vl_1}\right) \nonumber\\
& \times \vl_2 \cdot (\vl - \vl_2) \vl_1 \cdot (\vl_2 - \vl_1) (\vl - \vl_2 + \vl_1) \cdot (\vl_2 - \vl_1) \vl_1 \cdot (\vl - \vl_2) C_{l_1}^{EE} C_{|\vl - \vl_2|}^{\phi\phi} C_{|\vl_2 - \vl_1|}^{\phi\phi} \, .
\label{eq:tempxtempV}
\end{align}
The first, $U_l$, arises from correlating the (large-scale) lenses in each $\tilde{E}^{(1)}$; it is expected to dominate over $V_l$, which arises from correlations between the large- and small-scale lenses.

\begin{figure}
\centering
\includegraphics[width=0.8\textwidth]{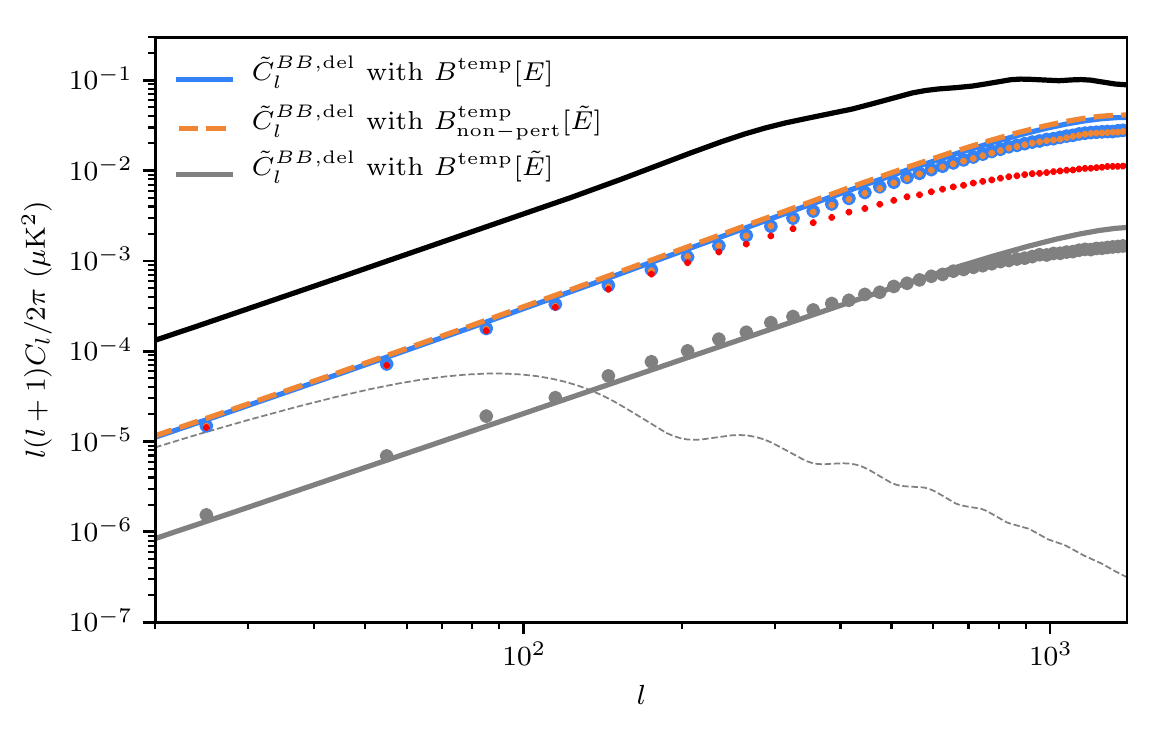}
\caption{Analytic (solid lines) versus simulated (dots) lensing $B$-mode residual spectra after delensing by: (i) building a gradient-order template from unlensed (blue) or lensed (grey) $E$-modes; and (ii) forming a non-perturbative template involving lensed $E$-modes (orange). We also show the simulated residual $B$-mode power after anti-lensing polarization fields containing lensed $E$- and $B$-modes (red dots). Noiseless polarization fields and perfect $\phi$ are used in all cases. Also shown is the lensing $B$-mode power spectrum (black, solid line) and a primordial $B$-mode spectrum for $r=0.001$ (dashed) generated using~\texttt{CAMB}~\cite{Lewis:1999bs}.}
\label{fig:sim_vs_analytic_both_cases}
\end{figure}

As expected from the similarity of the two fields highlighted in Fig.~\ref{fig:compare_lensB2_lensE1convphi}, an explicit numerical evaluation of all these terms\footnote{In order to deal appropriately with the highly oscillating, multi-dimensional integrands at hand, we employ a Monte-Carlo approach combining importance sampling with globally-adaptive subdivision of the integration domain, as implemented in the code \texttt{Suave}, part of the publicly-available \texttt{Cuba} library~\cite{cuba}.} reveals that $\langle \Delta \tilde{B}^{\text{temp}}(\vl) \Delta \tilde{B}^{\text{temp}}(\vl') \rangle'\approx \langle \tilde{B}^{(2)}(\vl) \tilde{B}^{(2)}(\vl') \rangle' \approx \langle \tilde{B}^{(2)}(\vl) \Delta \tilde{B}^{\text{temp}}(\vl') \rangle'$ (with $V_l \ll U_l$ by approximately two orders of magnitude). Consequently, there is extensive cancellation between leading-order contributions, and \emph{the residual lensing $B$-mode power spectrum after delensing with a gradient-order template built from lensed $E$-modes has an amplitude of $O(1)\,\%$ of the lensing $B$-mode power spectrum};  see Fig.~\ref{fig:sim_vs_analytic_both_cases}. This is in good agreement with simulations\footnote{We use 200 flat-sky simulations on a square grid with 1024 pixels per side and a pixel width of 1\,arcmin. To do this, we use the publicly-available code \texttt{QuickLens} (\url{https://github.com/dhanson/quicklens}, though an amended and extended version can be found at \url{https://github.com/abaleato/Quicklens-with-fixes}).} on large angular scales up to the accuracy afforded by our numerical integration and field remapping codes. On small scales, $l>1000$, the simulated and analytic spectra diverge somewhat, likely due to the significance of higher-order contributions missing from the analytic expressions. 

We can also recover the approximate cancellation between the terms in Eq.~\eqref{eq:BBresidualpower} by approximating the integrands in Eqs.~(\ref{eq:B2xtemp}--\ref{eq:tempxtempU}). The first of these is dominated by large-scale lenses in the construction of $\tilde{E}^{(1)}$ correlating with the large-scale lens in $\tilde{B}^{(2)}$, and the smaller-scale lens remapping $\tilde{E}^{(1)}$ to the large-scale $\Delta \tilde{B}^{\text{temp}}$ correlating with the small-scale lens in $\tilde{B}^{(2)}$. This corresponds to $|\vl_2 - \vl_1| \ll l_1$ in the integrand of Eq.~\eqref{eq:B2xtemp}. In the large-scale limit, we therefore have
\begin{align}
\langle \tilde{B}^{(2)}(\vl) \Delta \tilde{B}^{\text{temp}}(\vl') \rangle' &\approx  \int \frac{d^2 \vl_1}{(2\pi)^2}\,
\sin^2 2\left(\psi_{\vl_1}-\psi_{\vl}\right) \left[\vl_1 \cdot (\vl_1 - \vl)\right]^2 C_{l_1}^{EE} C_{|\vl - \vl_1|}^{\phi\phi}
\int \frac{d^2 \vl_2}{(2\pi)^2}\, \left[\vl_1\cdot(\vl_2-\vl_1)\right]^2 C_{|\vl_2 - \vl_1|}^{\phi\phi} \nonumber \\
&\approx \frac{1}{2} \langle \valpha^2 \rangle \int \frac{d^2 \vl_1}{(2\pi)^2}\,
\sin^2 2\left(\psi_{\vl_1}-\psi_{\vl}\right) l_1^6 C_{l_1}^{EE} C_{l_1}^{\phi\phi} \nonumber \\
&\approx \langle \tilde{B}^{(2)}(\vl)\tilde{B}^{(2)}(\vl')\rangle' \, .
\end{align}
Similarly, $U_l$ is dominated by $|\vl_2 - \vl_1| \ll l_1$, and in the large-scale limit it is straightforward to show that
$U_l \approx \langle \tilde{B}^{(2)}(\vl)\tilde{B}^{(2)}(\vl')\rangle'$.

\begin{figure}
\centering
\includegraphics[width=0.87\textwidth]{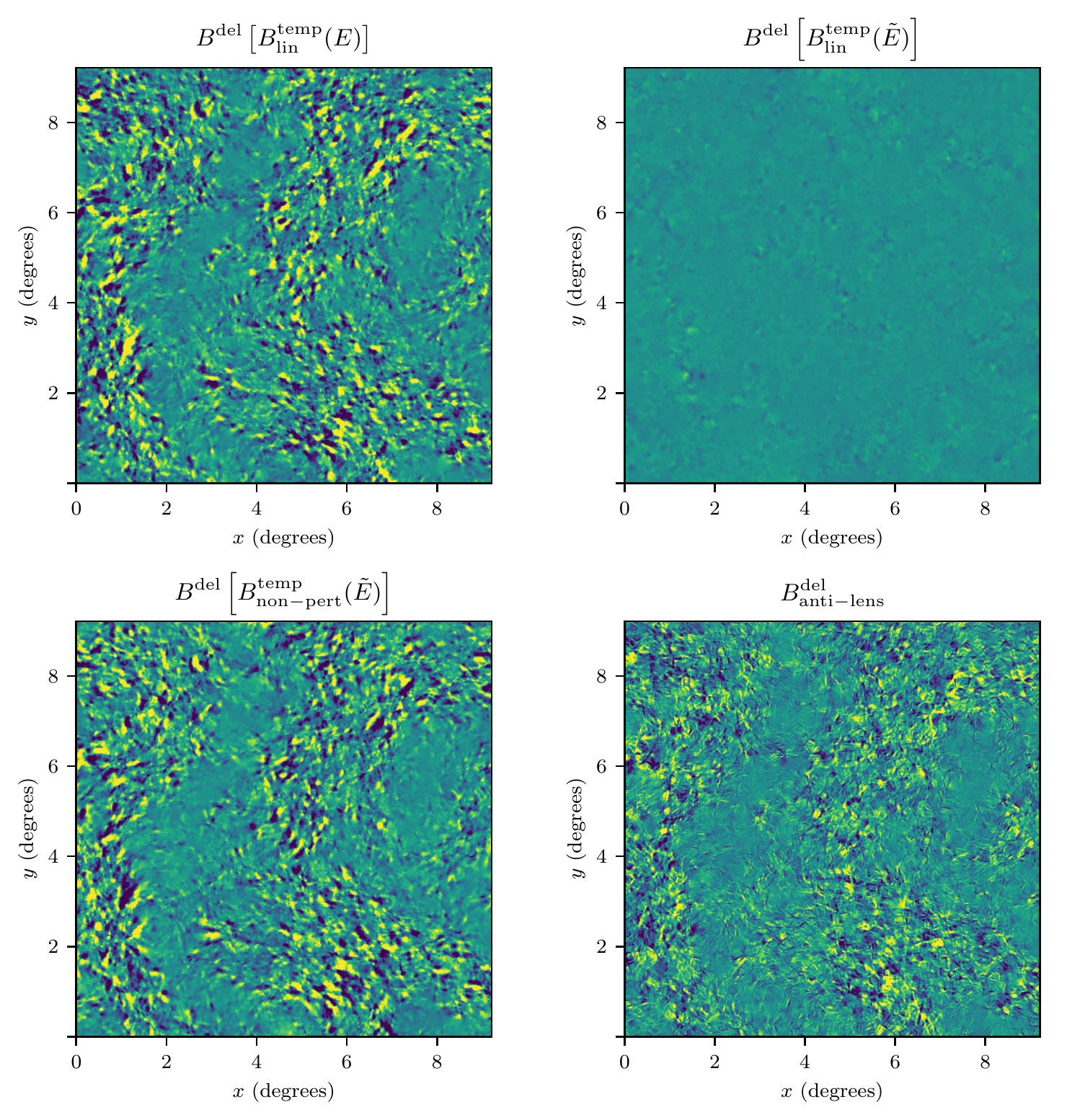}
\caption{Residual $B$-mode maps after delensing in the limit of noiseless CMB fields and access to the true $\phi$, and using either a gradient-order template built from unlensed $E$-modes (\emph{top left}), a gradient-order template built from lensed $E$-modes (\emph{top right}), a non-perturbative template built from lensed $E$-modes (\emph{bottom left}) or by anti-lensing polarization maps containing both $E$- and $B$-modes (\emph{bottom right}).The plotted intensity ranges from $-0.5\,\mu\text{K}$ (dark blue) to $0.5\,\mu\text{K}$ (yellow).}
\label{fig:residual_maps}
\end{figure}

For reference, the residual power obtained after delensing with a gradient-order template constructed from the lensed $E$-modes is one order of magnitude smaller than what can be attained by anti-lensing in the idealised case of perfect polarization observations and access to the true $\phi$; see Fig.~\ref{fig:sim_vs_analytic_both_cases} and Ref.~\cite{ref:diego-palazuelos_et_al_20}.
Anti-lensing attempts to remap the observed (possibly filtered) polarization field directly, approximating the inverse-remapping operation $P(\vx)=\tilde{P}(\vx+\valpha^{-1})$ with 
$P^{\mathrm{del}}(\vx) = \tilde{P}(\vx - \bm{\alpha})$, where $\valpha$ is evaluated at $\vx$. This incurs an error of $O(\bm{\alpha}\cdot\bm{\nabla}\bm{\alpha})$~\cite{Anderes:2014foa} since
\begin{align}
P^{\mathrm{del}}(\vx) &= P^E\bigl(\vx - \valpha + \valpha(\vx - \valpha)\bigr) \nonumber \\
&\approx P^E(\vx - \valpha\cdot \vnabla \valpha) \nonumber \\
&\approx P^E(\vx) - (\valpha\cdot \vnabla \valpha)\cdot \vnabla P^E(\vx) ,
\end{align}
where, in the first line, $\valpha(\vx - \valpha)$ denotes $\valpha$ evaluated at $\vx - \valpha(\vx)$. Extracting the $B$-modes gives
\begin{equation}
B^{\text{del}}(\vl) \approx - \mathcal{B}_\vl\left[(\valpha\cdot \vnabla \valpha)\cdot \vnabla P\right] = - \frac{1}{2} \mathcal{B}_\vl\left[ \left(\vnabla \valpha^2 \right)\cdot \vnabla P^E \right] ,
\end{equation}
where the relation $\valpha\cdot\vnabla \valpha = \vnabla \valpha^2 /2$ follows from $\valpha$ being a gradient. We see that the leading-order $B$-mode residuals after anti-lensing are the same as the $B$-modes produced by lensing $E$ with a lensing potential $-\valpha^2/2$. Simulated residuals after anti-lensing noise-free polarization with the true $\phi$ are shown in Fig.~\ref{fig:residual_maps}, along with the residuals for template delensing.
The power spectrum of the anti-lensing residuals on large scales is very similar to the power spectrum after gradient-order template delensing with the unlensed $E$-modes (i.e., approximately the power spectrum of $\tilde{B}^{(2)}$); see Fig.~\ref{fig:sim_vs_analytic_both_cases}.
This is because
\begin{equation}
- \frac{1}{2} \left(\vnabla \valpha^2 \right)\cdot \vnabla P^E = -\frac{1}{2} \vnabla \cdot \left(\valpha^2 \vnabla P^E\right) + \frac{1}{2} \valpha^2 \nabla^2 P^E ,   
\end{equation}
and the first term on the right is a total divergence, which is suppressed on large scales compared to the second term. The power spectrum of $\mathcal{B}_\vl[\valpha^2 \nabla^2 P^E]/2$ evaluates to
\begin{equation}
\frac{1}{4}\langle \mathcal{B}_\vl[\valpha^2 \nabla^2 P^E] \mathcal{B}_{\vl'}[\valpha^2 \nabla^2 P^E]\rangle' = \frac{1}{2} \int \frac{d^2
  \vl_1}{(2\pi)^2} \int \frac{d^2 \vl_2}{(2\pi)^2} \, \sin^2
2(\psi_{\vl_1} - \psi_{\vl} ) l_1^4 \left[\vl_2\cdot
  (\vl-\vl_1-\vl_2)\right]^2 C_{l_1}^{EE} C_{l_2}^{\phi\phi}
C_{|\vl-\vl_1 - \vl_2|}^{\phi\phi} \, ,
\end{equation}
the same as for $\tilde{B}^{(2)}$, Eq.~\eqref{eq:twotwo}, but with $(\vl_1\cdot \vl_2)^2[\vl_1\cdot (\vl - \vl_1 - \vl_2)]^2$ replaced by $l_1^4 [\vl_2 \cdot (\vl - \vl_1 - \vl_2)]^2$. However, both of these geometric couplings become equal in the limit $l \ll l_1$ and $l_2 \ll l_1$ (and the disjoint region with $\vl_2 \rightarrow \vl - \vl_1 - \vl_2$), which dominate the integral on large scales.

We end this section by noting a further apparent benefit of delensing with a gradient-order template constructed from the lensed $E$-modes. When adjusted to an appropriate color scale, the residuals shown in Fig.~\ref{fig:residual_maps} for the noise-free case appear to the eye to be significantly more Gaussian than for either anti-lensing, a gradient-order template made from the unlensed $E$-modes, or a non-perturbative template with the lensed $E$-modes (which we discuss in detail in the next section). In the future, when noise levels permit very aggressive delensing, having more Gaussian residuals may simplify the subsequent likelihood analysis, e.g., by reducing the covariance between power spectrum estimates at different scales. We defer a more quantitative analysis of the statistics of these residuals to future work.

\section{Delensing with a non-perturbative template}
\label{sec:delensed_psnp}

We now consider the case of $B$-mode delensing with a template constructed from the lensed $E$-modes but with the lens remapping handled non-perturbatively (Eq.~\ref{eq:template_np}). This was the approach adopted in Refs.~\cite{Planck2018:lensing, ref:polarbear_delensing_19}, for example.
As we shall see, this case performs similarly badly in the noise-free limit to the gradient-order template constructed from unlensed $E$-modes. 

Expanding the non-perturbative template to second order in $\phi$, we have
\begin{align}
\tilde{B}^{\text{temp}}_{\text{non-pert}}(\vl) &= \mathcal{B}_\vl\left(\alpha^i \nabla_i P^E\right) + \mathcal{B}_\vl\left(\alpha^i \nabla_i P^{\tilde{E}^{(1)}}\right) + \frac{1}{2} \mathcal{B}_\vl\left(\alpha^i \alpha^j \nabla_i \nabla_j P^{E}\right) + O(\phi^3) \nonumber \\
&= \tilde{B}^{(1)}(\vl) + \Delta \tilde{B}^{\text{temp}}(\vl) + \tilde{B}^{(2)}(\vl) + O(\phi^3) \, .
\label{eq:Btempnp}
\end{align}
It follows that the leading-order residuals after delensing with such a template are $-\Delta \tilde{B}^{\text{temp}}(\vl)$. Notice now how the term $\tilde{B}^{(2)}(\vl)$ is absent from the residuals, in contrast to the case of the gradient-order template. It follows that the residual power after delensing with the non-perturbative template is $O(10)\,\%$ of the original lensing power in the noise-free limit, reintroducing this delensing floor; see Fig.~\ref{fig:sim_vs_analytic_both_cases}. Furthermore, the residuals, a simulation of which are shown in Fig.~\ref{fig:residual_maps}, appear to the eye to be less Gaussian for the non-perturbative template than those obtained for the gradient-order template. \emph{We see that for templates constructed directly from the observed $E$-modes, it is preferable to construct the template at gradient order rather than with non-perturbative remapping.} In the case of non-perturbative templates built from anti-lensed $E$-modes, there would be no delensing floor of this nature, since anti-lensing removes $\tilde{E}^{(1)}$ and consequently $\Delta \tilde{B}^{\text{temp}}(\vl)$ would be absent from Eq.~\eqref{eq:Btempnp}\footnote{This is in agreement with what was seen in simulations by Ref.~\cite{ref:diego-palazuelos_et_al_20}.}.

So far we have considered the idealised case of noiseless $E$-modes and access to the true $\phi$.
We now demonstrate that the conclusion that a gradient-order template is preferable over a non-perturbative one still holds in the practical case of noisy $E$-modes and a lensing proxy $\phi^{\text{proxy}}$ that is only partially correlated with the true $\phi$, albeit with more marginal benefits. The residuals after delensing are more complicated in this case, in part because the term that is first-order in $\phi$ no longer vanishes. The gradient-order and non-perturbative templates share the same $O(\phi)$ residuals
\begin{equation}
\tilde{B}^{\text{temp}}_{\text{res}}(\vl) = - \int \frac{d^2 \vl_1}{2\pi}\, \sin 2(\psi_{\vl_1} -
\psi_{\vl}) \vl_1 \cdot (\vl-\vl_1) \Bigl(E(\vl_1)\phi(\vl-\vl_1)-\calW^E_{l_1}
[E(\vl_1)+n^E(\vl_1)]
\calW^\psi_{|\vl-\vl_1|}\phi^{\text{proxy}}(\vl-\vl_1)\Bigr) \, ,
\end{equation}
where $n^E$ is the noise on the observed $E$-modes.
Indeed, it is the power spectrum of this term that is usually assumed to dominate the $B$-mode signal power after delensing:
\begin{equation}\label{eqn:delensed_power}
        \tilde{C}_l^{BB,\mathrm{res}} = \int \frac{d^2 \vl_1}{(2\pi)^2} \left[\vl_1 \cdot (\vl-\vl_1)\,\sin 2(\psi_{\vl_1} - \psi_\vl)\right]^2 C_{l_1}^{EE}C_{|\vl-\vl_1|}^{\phi\phi} \left(1- \mathcal{W}^{E}_{l_1}\calW^{\phi}_{|\vl-\vl_1|} \right) \, .
    \end{equation}
For sufficiently-low $E$-mode noise levels and a highly-correlated lensing proxy, the residual power after delensing may instead be dominated by terms that are second order in $C_l^{\phi\phi}$, particularly for the case of a non-perturbative template. Now, there will be contributions from both ``2--2'' and ``1--3'' terms (which may partially cancel, as for the lensed $B$-mode spectrum), although as the ideal limit is approached the former will dominate.

Although the residual power at second order in $C_l^{\phi\phi}$ could be calculated along similar lines to the calculations for the ideal cases in Sec.~\ref{sec:delensed_ps}, instead we shall estimate the residual power from simulations. We simulate the observed, lensed $E$-modes by remapping a realization of unlensed $E$-modes with a Gaussian realization of $\phi$, and subsequently add white noise with variance $\Delta_P^2$. For the lensing proxy, we add white noise to the lensing convergence ($\kappa = -\nabla^2 \phi /2$), with variance $\Delta_\kappa^2$. This mimics the statistical noise that arises when reconstructing $\phi$ internally from the CMB\footnote{We choose not to use the usual quadratic estimator noise power because this would detract from the pedagogical value of this exercise without necessarily making it more realistic. The reason is that quadratic estimators will be superseded by iterative delensing before the lower range of $\kappa$ sensitivities considered here --- those that unveil the delensing floors --- can be reached. For higher noise levels and instrument resolutions of $O(1)\,\mathrm{arcmin}$, we find the quadratic estimator noise power for $\kappa$ to be quite well approximated as white noise on the relevant scales, such that the different treatments produce only percent-level differences on the spectrum of delensed $B$-modes on scales  $l<300$.} (e.g., Ref.~\cite{Hu:2001tn}). Such reconstruction noise is approximately white on large scales, and generally remains so on all scales where the reconstruction is signal dominated. For reference, white noise with (constant) power equal to the peak power in the convergence power spectrum, $C_l^{\kappa \kappa} \approx 2.2\times 10^{-7}$ at $l\approx 30$, corresponds to $\Delta_\kappa = 1\,\text{arcmin}$ (i.e., the standard deviation of the white noise averaged in a pixel of side $1.6\,\text{arcmin}$ is $1.6$). In addition, the large-scale white noise level for a (non-iterative) reconstruction from a CMB survey with $\Delta_P = 1\,\mu\text{K\,arcmin}$ (and using multipoles up to $l=3000$) is $0.26\,\text{arcmin}$.
Gradient-order and non-perturbative templates are constructed from the Wiener-filtered observed $E$-modes and lensing proxy, and delensed fields are obtained by subtracting these from simulated, noiseless lensed $B$-modes. In this way, we can isolate the change in $B$-mode power due to delensing. 

\begin{figure}
\centering
\includegraphics[width=0.9\textwidth]{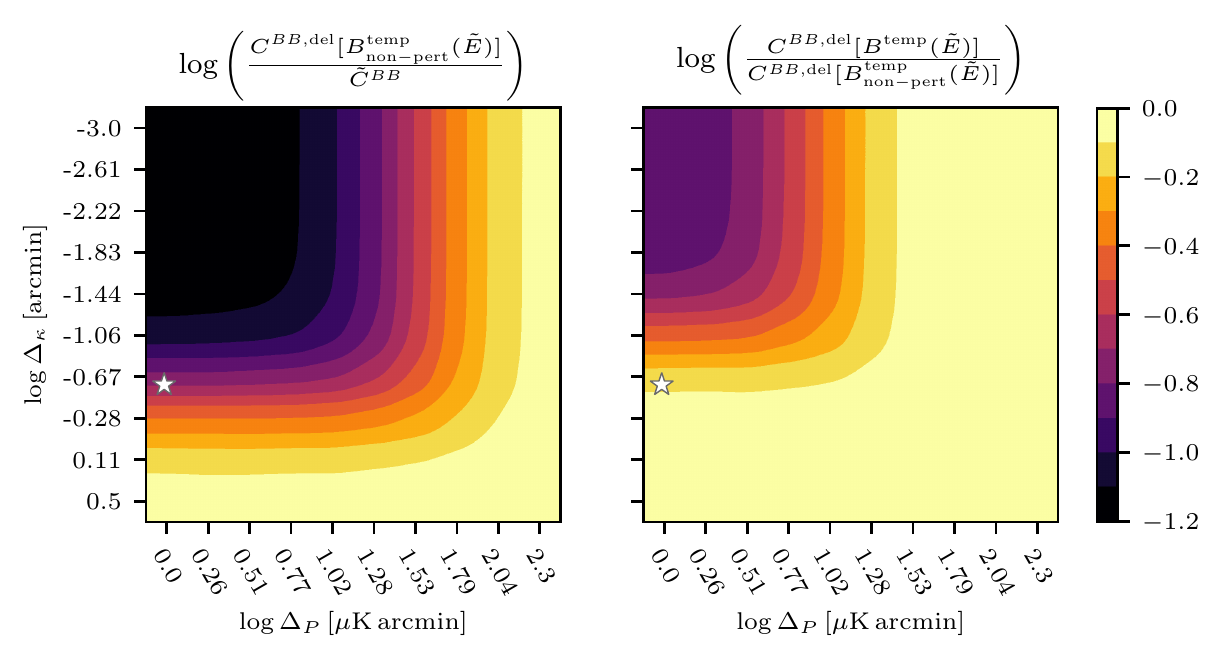}
\caption{\emph{Left}: Fraction of lensing power remaining after delensing noiseless $B$-modes with a non-perturbative template built from (Wiener-filtered) noisy, lensed $E$-modes and noisy $\phi$. \emph{Right}: Ratio of residual power after delensing with a gradient-order template to that after delensing with a non-perturbative template, in both cases using filtered, noisy, lensed $E$-modes and $\phi$ to delens noiseless $B$-modes. When computing the ratios, we first average spectra over the multipole range $l\in[25,325]$, and over 20 simulations. In all cases, we add white noise to the simulated lensed $E$-modes used to construct the templates with variance $\Delta_P^2$ (and ignore the effects of finite instrumental angular resolution). For the lensing proxy, we add white noise to the true lensing convergence ($\kappa = -\nabla^2 \phi /2$) with variance $\Delta_\kappa^2$ per $\text{arcmin}^2$ pixel. Note the $O(10)\,\%$ floor in delensed power when using the non-perturbative template in the ideal limit (top-left corner), and how the gradient-order template consistently outperforms the non-perturbative one, even before the latter has hit its delensing floor. The improvement will start to be significant for non-iterative lensing reconstructions in the era of CMB-S4~\cite{ref:s4_science_book} (marked with stars).}
\label{fig:comparing_templates_with_filtering}
\end{figure}

The results of these simulations are given in Fig.~\ref{fig:comparing_templates_with_filtering}. The left panel shows the ratio of the delensed power using the non-perturbative template to the original lensing $B$-mode power. For sufficiently low noise levels (top-left corner), we recover the $O(10)\,\%$ floor in the delensed power that we uncovered in the idealised case. The right panel shows the ratio of the residual powers after delensing with the gradient-order template and the non-perturbative template. The former gives lower residual power for all noise levels. As the ideal limit is approached, we recover the ratio of $O(0.1)$ seen earlier for the ideal case. In the opposite limit, where noise is significant in either the observed $E$-modes or the lensing proxy, the ratio tends to one with the delensed power given by Eq.~\eqref{eqn:delensed_power} for both templates. For the specifications of a nominal Stage-4 CMB experiment ($\Delta_{P} = 1\,\mu\mathrm{K}\,\mathrm{arcmin}$ and $\Delta_\kappa = 0.26\,\text{arcmin}$)~\cite{ref:s4_science_book}, the residual power after gradient-order delensing is approximately $80\,\%$ of that obtained using a non-perturbative template. This translates to the removal of an additional $4.5\,\%$ of the lensing power originally in the maps.

\section{Conclusions}

The variance associated with lensing $B$-modes ought to be mitigated in order to optimise searches for the primordial signal associated with a stochastic background of gravitational waves that may have been produced during cosmic inflation. The partial removal of lensing, known as delensing, is often carried out by combining high-resolution $E$-mode observations with some proxy of the lensing potential to produce a template-estimate of the lensing $B$-modes, which can then be subtracted from observations on large angular scales. This was the approach followed by the \emph{Planck}~\cite{Planck2018:lensing}, SPT~\cite{ref:spt_17} and POLARBEAR~\cite{ref:polarbear_delensing_19} collaborations in their successful demonstrations of delensing on real $B$-mode data.

In this paper, we have considered the limitations of this template method that arise from lensing of the $E$-modes used in the template and the remapping approximations made in its construction.
Gradient-order templates, where the lensing action on the $E$-mode observations is approximated by the gradient term in a Taylor expansion, are often used, particularly in forecasting work given their analytic simplicity. Such templates are usually presumed to be highly accurate in tracking the true lensing $B$-modes if the polarization measurements are of sufficient precision and an estimate of the lensing deflections is available that is highly correlated with the true lensing. This accuracy is often assumed to follow from the fact that higher-order corrections to the leading-order calculation of the lensing $B$-mode spectrum are very small, at the $O(1)\,\%$ level. However, we showed that in the case of the lensed $B$-mode spectrum, the small contribution from terms beyond gradient order is a result of cancellations between terms that are separately relatively large (around $10\,\%$), and that these cancellations are not necessarily relevant for the delensed spectrum. In particular, a gradient-order template constructed from the unlensed (or, more realistically, the delensed or anti-lensed) $E$-modes introduces a floor in the residual power of around 10\,\% of the original lensing $B$-mode power due to unsubtracted terms in the lensing $B$-modes that are second order in the lensing deflections. Fortunately, in the case of a gradient-order template constructed from the lensed $E$-modes, cancellations appear at higher order which reduce the floor in the residual power to the $O(1)\,\%$ level. The dominant cancellations in this case are between the second-order term in the lensing $B$-modes and that in the template arising from the first-order lensing correction to the $E$-modes. The larger $10\,\%$ residual-power floor also arises in anti-lensing, where one displaces the full lensed polarization field using minus the lensing deflections as an approximation to the true inverse remapping.

Non-perturbative templates are also sometimes considered, in which a template is made by directly deflecting the (filtered) observed $E$-modes by the lensing proxy, rather than relying on the gradient-order approximation. Importantly, we showed that such templates are also fundamentally limited, reintroducing a floor of $O(10)\,\%$ in the delensed power. This behaviour arises since the first-order lensing correction to the $E$-modes used in the template are no longer approximately cancelled. We further showed that in practical applications of delensing, where noisy $E$-modes and a partially-correlated lensing proxy are used, the benefits of the gradient-order template persist, albeit with more marginal gains. Indeed, the better performance of the gradient-order template become significant well before the non-perturbative template has hit its delensing floor: for an experiment with characteristics similar to CMB-S4, this would enable the removal of an additional $5\,\%$ of lensing power, reducing the lensing-related uncertainty on the tensor-to-scalar ratio $r$ by a factor of around 1.2 in the limit $r=0$. We therefore recommend that in practical applications of $B$-mode template delensing, where the template is constructed directly from the (filtered) observed $E$-modes, the gradient-order approach should be used rather than a non-perturbative remapping.

\section*{Acknowledgments}
ABL gratefully acknowledges support from an Isaac Newton Studentship at the University of Cambridge and the Science and Technology Facilities Council (STFC). 
AC acknowledges support from the STFC (grant numbers ST/N000927/1 and ST/S000623/1). JC acknowledges support from a SNSF Eccellenza Professorial Fellowship (No. 186879).
The authors would like to thank Toshiya Namikawa for useful discussions at the early stages of this project.
This work made use of \texttt{Mathematica}~\cite{Mathematica}, \texttt{NumPy}~\cite{numpy} and \texttt{Matplotlib}~\cite{matplotlib}.
\bibliography{lensing.bib}

\end{document}